\documentclass[12pt,a4paper,notitlepage]{article}

\usepackage{amssymb,amsmath}
\usepackage[dvips]{graphicx}

\newcommand{\bsl}{\boldsymbol}

\begin{document}

\author{I.M.Narodetskii\thanks{e-mail: naro@heron.itep.ru},
Yu.A.Simonov\thanks{e-mail: simonov@heron.itep.ru},
M.A.Trusov\thanks{e-mail: trusov@heron.itep.ru}, and
A.I.Veselov\thanks{e-mail: veselov@heron.itep.ru} \\[4mm]
\textit{Institute of Theoretical and Experimental Physics} \\[2mm]
\textit{Russia, 117218, Moscow, B. Cheremushkinskaya str., 25}}

\title{Pentaquark spectrum in string dynamics}

\date{}

\maketitle

\begin{abstract}
\noindent The masses of $uudd\bar s $ and $uudd\bar d$ pentaquarks
are evaluated in a framework of  the Effective Hamiltonian
approach to QCD using the Jaffe-Wilczek $[ud]^2\bar q$
approximation. The mass of the  $[ud]^2\bar s$ state is found to
be $\sim 400$ MeV higher than the observed $\Theta^+(1540)$ mass.
\end{abstract}

Recently several collaborations reported the observation of a very
narrow peak in the $K^+N$ invariant mass distribution
\cite{Nakano:2003}-\cite{Dolgolenko}. The reported masses have
been clustered around $1540$ MeV, with widths less than $25$ MeV.
These results are in remarkable agreement with a chiral soliton
model prediction \cite{Diakonov:1997} of a $\Theta^+$ state at
$1530$ MeV and a width less than $15$ MeV \footnote{The problem of
the $\Theta^+$ width is crucial since the widths beyond the
few-MeV level seem to be excluded by the $K^+N$ phase-shift
analysis \cite{Arndt} and total cross sections \cite{Nussinov}.}.

In the sense of the quark model such a state is manifestly exotic
-- it is not a simple three-quark state but rather the pentaquark
$uudd\bar s$ state. If so,  the discovery of $\Theta^+$ provides
an opportunity to refine our understanding of nonperturbative
quark dynamics at low energy. The $\Theta$-hyperon has hypercharge
$Y=2$ and third component of isospin $I_3=0$. The apparent absence
of the $I_{3}=+1$, $\Theta^{++}$ in $K^{+}p$ argues against $I=1$,
therefore it is usually assumed the $\Theta$ to be an isosinglet,
although other suggestions have been made in the literature
\cite{capstick-page}. The other quantum numbers are not
established yet. The uncorrelated quark models, in which all
quarks are in the same spatial wave function, naturally predict
the ground state energy of a $J^P=\frac{1}{2}^-$  pentaquark to be
lower than that of a $J^P=\frac{1}{2}^+$ one. Several suggestions
were made to reverse this order \cite{Jaffe:2003}, \cite{stancu}
but no quantitative microscopic evaluations have been performed
yet. The QCD sum rules predict a negative parity $\Theta^+$ of
mass $\simeq 1.5$ GeV, while no positive parity state was
found~\cite{sumrules}. The lattice QCD study also predicts that
the parity of the $\Theta(1540)$ is most likely negative
\cite{Fodor},\cite{Sasaki}.

Jaffe and Wilczek proposed \cite{Jaffe:2003} that for
$\Theta(1540)$ with $I^P=0^+$ and other $q^{4}\bar q$ baryons the
four quarks are bound into two scalar, singlet isospin  diquarks
combined with the antiquark into a color singlet. The problem of
diquarks has a long history  and sound physical motivation
\cite{anselmino}. Both gluon and pion exchanges in the singlet
spin and isospin $ud$ state result in the strong short-range
attraction  which may keep $u$ and $d$ quarks close together thus
forming almost point-like diquark.

The purpose of this letter is to test this interpretation
quantitatively using the Effective Hamiltonian (EH) approach in
QCD \cite{lisbon}. An attractive feature of this formalism is that
it contains the minimal number of input parameters: current (or
pole) quark masses, the string tension $\sigma$ and the strong
coupling constant $\alpha_s$, and does not contain fitting
parameters as e.g. the total subtraction constant in the
Hamiltonian. Therefore the EH method is the best way to compute
the masses of yet unknown quark systems like pentaquarks.

\begin{table}
\caption{Comparison of the EH approach predictions for
spin--averaged masses of nucleons, hyperons, and pentaquark with
lattice results and experimental data (assuming the \(\Theta^+\)
quantum numbers are \(I^P=0^+\)). The masses are given in units of
GeV.}
\begin{center}
\begin{tabular}{|l|l|l|l|l|}
\hline & \((N,\Delta)\) & \((N^-,\Delta^-)\) &
\((\Lambda,\Sigma,\Sigma^*)\) & \(\Theta^+\)~(\(I^P=0^+\)) \\
\hline Lattice & 1.07~~\cite{CP-PACS} & 1.76~~\cite{UKQCD} &
1.21~~\cite{CP-PACS} & 2.80~~\cite{Fodor} \\ \hline EH &
1.14~~\cite{plekhanov} & 1.63~~\cite{baryons} &
1.24~~\cite{plekhanov} & $>2.12$ \footnotemark
\\ \hline Experiment & 1.08 & 1.62 & 1.27 & 1.54 \\ \hline
\end{tabular}
\end{center}
\label{table_compare}
\end{table}
\footnotetext{This work}

In order to illustrate the accuracy of the method, we quote in
Table \ref{table_compare} a few EH results for baryons compared
with the lattice ones and experiment. One can observe that the
accuracy of the EH method for the three-quark systems is \(\sim
100\) MeV or better. We expect the same accuracy for the
diquark-diquark-(anti)quark system considered in
\cite{Jaffe:2003}.

The EH for the three constituents has the form
\begin{equation}
\label{EH} H=\sum\limits_{i=1}^3\left(\frac{m_i^{2}}{2\mu_i}+
\frac{\mu_i}{2}\right)+H_0+V,
\end{equation}
where $H_0$ is the kinetic energy operator, $V$ is the sum of the
perturbative one-gluon exchange potentials and the string
potential which is proportional to the total length of the string,
{\it i.e} to the sum of the  distances of (anti)quark or diquarks
from the string junction point. The dynamical masses $\mu_i$ are
expressed in terms of the current masses $m^{}_i$ from the
condition of the minimum of the hadron mass $M_H^{(0)}$ as
function of $\mu_i$ \footnote{Technically, this is done using the
auxiliary field approach to get rid of the square root term in the
Lagrangian \cite{polyakov}, \cite{brin77}. Applied to the QCD
Lagrangian, this technique yields the EH  for mesons or baryons
depending on auxiliary fields $\mu_i$. In practice, these fields
are finally treated as $c$-numbers determined from
(\ref{minimum_condition}).}:
\begin{equation} \label{minimum_condition}
\frac{\partial M_H^{(0)}(m_i,\mu_i)}{\partial \mu_i}=0, ~~~
M_H^{(0)}=\sum\limits_{i=1}^3\left(\frac{m_i^{2}}{2\mu_i}+
\frac{\mu_i}{2}\right)+E_0(\mu_i), \end{equation} $E_0(\mu_i)$
being eigenvalue of the operator $H_0+V$. The physical mass $M_H$
of a hadron is
\begin{equation}\label{self_energy}
M_H=M_H^{(0)}+\sum_i C_i,
\end{equation}
where the constants $C_i$ have the meaning of the constituents
self energies and are expressed in terms of string tension
$\sigma$ \cite{simonov_self_energy}. The self-energy is due to
constituent spin interaction with the vacuum background fields and
equals zero for any scalar diquark.

\begin{figure}
\begin{center}
\begin{tabular}{ccc}
\includegraphics[width=60mm,keepaspectratio=true]{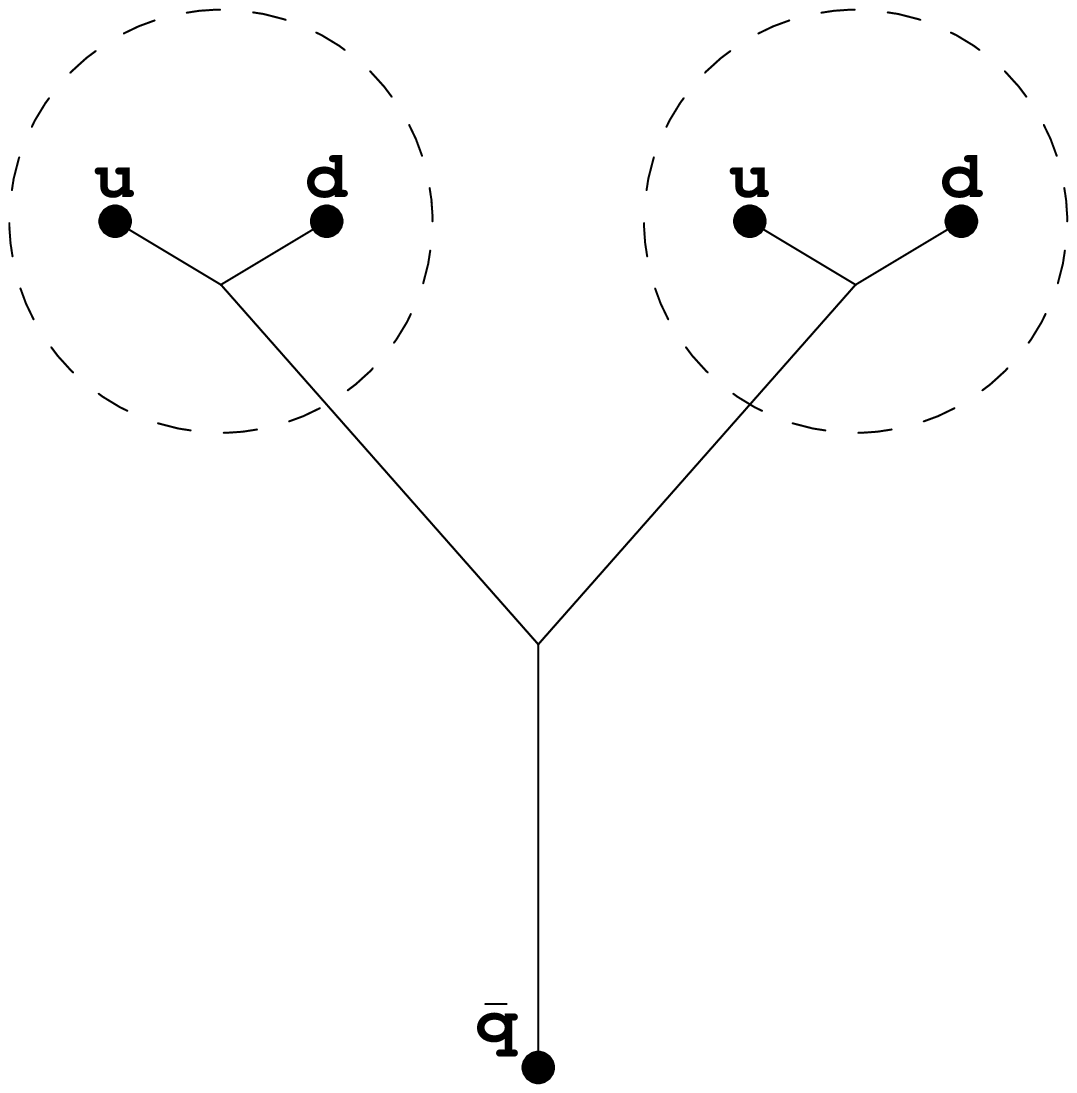} &
\hphantom{aaaaa} &
\includegraphics[width=50mm,keepaspectratio=true]{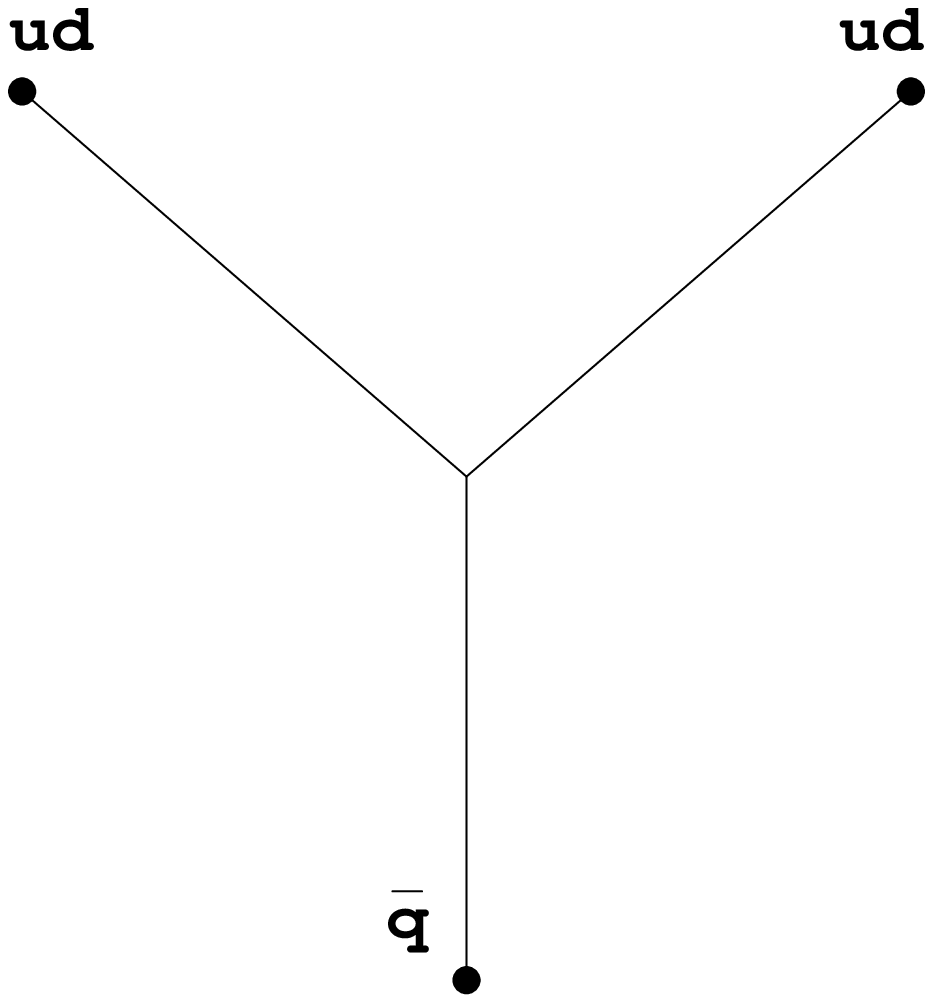} \\
{\Large $a$} & & {\Large $b$}
\end{tabular}
\end{center}
\begin{quote}
\caption{The Jaffe-Wilczek reduction of the $uudd\bar q$
pentaquark $(a)$ to the effective $[ud]^2\bar q$ problem $(b)$}
\label{fig}\end{quote}
\end{figure}

In the quark model  five quarks are connected by seven strings
(Fig. 1a). In the diquark approximation the short legs on this
figure shrink to points and the five-quark system effectively
reduces to the three-body one (Fig. 1b), studied within the EH
approach in \cite{baryons},\cite{trusov}. Consider a pentaquark
consisting of two identical diquarks with current mass $m_{[ud]}$
and antiquark with current mass $m_{\bar q}$ ($q=d,s$). The
three-body problem is conveniently solved in the hyperspherical
formalism \cite{fabre}. The wave function $\psi(
{\bsl\rho},{\bsl\lambda)}$ expressed in terms of the Jacobi
coordinates $\bsl \rho$ and $\bsl \lambda$ can be written in a
symbolical shorthand as
\begin{equation}\psi(\bsl{\rho},\bsl{\lambda})=\sum\limits_K\psi_K(R)Y_{[K]}(\Omega).
\end{equation}where $Y_{[K]}$ are eigen functions (the
hyperspherical harmonics) \cite{simonov_hyperspherical} of the
angular momentum operator $\hat K(\Omega)$ on the 6-dimensional
sphere: $\hat{K}^2(\Omega)Y_{[K]}=-K(K+4)Y_{[K]}$, with $K$ being
the grand orbital momentum. For \emph{identical} diquarks, like
$[ud]^{2}$, the lightest state must have a wave function
antisymmetric under diquark space exchange. There are two possible
pentaquark wave functions antisymmetric under diquark exchange,
one (with lower energy) corresponding to the total orbital
momentum $L=1$, and the second one (with higher energy)
corresponding to $L=0$. For a state with
$L=1,~l_{\rho}=1,~l_{\lambda}=0$ the wave function in the lowest
hyperspherical approximation $K=1$ reads
\begin{equation}
 \psi=R^{-5/2}\chi_1(R)u_1(\Omega),~~~
 u_1(\Omega)=\sqrt{\frac{8}{\pi}}\sin\theta\cdot
 (\bsl{n}_{\bsl{\rho}})_z,
\end{equation} where $R^2=\bsl{\rho}^2+\bsl{\lambda}^2$. For a state with
$L=0,~l_{\rho}=1,~l_{\lambda}=1$ the wave function  in the lowest
hyperspherical approximation $K=2$ is
\begin{equation}\psi=R^{-5/2}\chi_2(R)u_2(\Omega),~~~
u_2(\Omega)=\frac{4}{\sqrt{\pi^3}}\sin\theta\cos\theta\cdot
(\bsl{n}_{\bsl{\rho}} \bsl{n}_{\bsl{\lambda}}).\end{equation} The
Schr\"odinger equation written in terms of the variable
$x=\sqrt{\mu} R$, where $\mu$ is an arbitrary scale of mass
dimension which drops off in the final expressions, reads:
\begin{equation} \label{shr}
\frac{d^2\chi_K(x)}{dx^2}+
2\left[E_0+\frac{a_K}{x}-b_Kx-\frac{(K+\frac{3}{2})(K+\frac{5}{2})}{2x^2}\right]\chi_K(x)=0.,
\end{equation}
with the boundary condition $\chi_K(x) \sim {\cal O} (x^{5/2+K})$
as $x\to 0$ and the asymptotic behavior $\chi_K(x)\sim
{\mathrm{Ai}}((2b_K)^{1/3}x)$ as $x\to \infty$. In Eq. (\ref{shr})
\begin{equation}
\begin{aligned}
a_K&=R\sqrt{\mu}\cdot \int
V_{\text{Coulomb}}(\bsl{r}_1,\bsl{r}_2,\bsl{r}_3)\cdot u_K^2\cdot
d\Omega ,\\ b_K&=\frac{1}{R\sqrt{\mu}}\cdot\int
V_{\text{string}}(\bsl{r}_1,\bsl{r}_2,\bsl{r}_3)\cdot u_K^2\cdot
d\Omega ,
\end{aligned} \label{ab_int}
\end{equation}
where
\begin{equation}
V_{\text{Coulomb}}(\bsl{r}_1,\bsl{r}_2,\bsl{r}_3)=
-\frac{2}{3}\alpha_s\cdot\sum\limits_{i<j}\frac{1}{r_{ij}},
\end{equation}
and explicit expression of
$V_{\text{string}}(\bsl{r}_1,\bsl{r}_2,\bsl{r}_3)$ in terms of
Jacobi variables is given in \cite{plekhanov}. Note, that in the
Y-shape in Fig. $1b$, the strings meet at $120^\circ$ in order to
insure the minimum energy. This shape moves continuously to a
two--legs configuration where the legs meet at an angle larger
than $120^\circ$.

The mass of the $\Theta^+$ obviously depends on $m_{[ud]}$ and
$m_s$. The current masses of the light quarks are relatively
well-known: $m_{u,d}\approx 0$, $m_s\approx 170$ MeV. The
effective mass of the diquark $m_{[ud]}$ is basically unknown. In
principle, this mass could be computed dynamically. Instead, one
can estimate $m_{[ud]}$  from the nucleon mass calculations in the
quark-diquark approximation. In this case, neglecting for
simplicity Coulomb-like interaction, one obtains for the mass of
the nucleon: \begin{equation}
M_N=M_N^{(0)}-\frac{2\sigma}{\pi\mu_d}, \end{equation} where
$M_N^{(0)}$ and $\mu_d$ are defined from the minimum condition,
\begin{equation}
\frac{\partial M_N^{(0)}}{\partial \mu_{[ud]}}=\frac{\partial
M_N^{(0)}}{\partial \mu_d}=0. \label{min_cond}
\end{equation}
In Eq. (\ref{min_cond})
\begin{equation}
M_N^{(0)}=
\frac{m_{[ud]}^2}{2\mu_{[ud]}}+\frac{\mu_{[ud]}+\mu_d}{2}+E_0,\quad
E_0=\gamma\left(\frac{\sigma^2}{2}\cdot\frac{\mu_{[ud]}+\mu_d}{\mu_{[ud]}\mu_d}\right)^{1/3}
\end{equation}
with $\gamma=2.338$ being the first zero of the Airy function:
$\mathrm{Ai}(-\gamma)=0$. Equating $M_N$ to the experimental value
of the $N-\Delta$ center of gravity ($M_N=1.085$ GeV), we obtain
that $m_{[ud]}$ varies in the interval $0\le m_{[ud]}\le 300$ MeV,
when $\sigma$ varies in the interval
$0.15\text{~GeV}^2\le\sigma\le 0.17\text{~GeV}^2$. The last value
of $\sigma$ is preferred by meson spectroscopy, while the former
one follows from  the lattice calculations of Ref. \cite{TMNS02}.
The lowest $uudd\bar s$ pentaquark  corresponds to the case when
one fixes the effective mass of a diquark $m_{[ud]}=0$. In what
follows, we use $m_{[ud]}=0$, $\sigma=0.15\text{~GeV}^2$, but
explicitly include the Coulomb--like interaction between quark and
diquarks.

The calculated masses of $[ud]^2\bar s$   pentaquarks are given in
Table \ref{table_masses}. For illustration we show also the masses
of $[ud]^2\bar d$ pentaquarks which have been identified in
\cite{Jaffe:2003} with the otherwise perplexing Roper resonance.
The states with $K=2$ are always higher than those with $K=1$ by
$\sim$ 300 MeV, the difference being mainly due to the kinetic
energy term in (\ref{shr}). Increasing $\alpha_s$ up to $0.6$ (the
value used in the Capstick-Isgur model \cite{capstick-isgur})
decreases the $[ud]^2\bar s$ mass by $\sim$ $100$ MeV. The
hyperfine interaction due to $\eta$ exchange between diquarks and
strange antiquark can lower the $\Theta^+$ energy by $\sim 100$
MeV or less \cite{stancu}. Recall that the accuracy of our
approach for pentaquarks is expected to be $\sim 100$ MeV. Since
the calculated value is the lower bound on the pentaquark mass, we
conclude that the lightest $uudd\bar s$ state is still $\sim 400$
MeV higher than the observed $\Theta^+(1540)$ state. Note that
$[ud]^2\bar d$ pentaquarks lie $\sim 100$ MeV lower than
$[ud]^2\bar s$ pentaquarks. This is the consequence of the
different current masses $m_{\bar d}$ and $m_{\bar s}$. Also note,
that for the \(\frac{1}{2}^-\) pentaquark the \([ud][d\bar{s}]u\)
configuration with \(K_{\text{min}}=0\) is more preferable giving
the mass \(\gtrsim 1.8\) GeV.

\begin{table}
\caption{The pentaquark results in the  quark-diquark-diquark
approximations. Shown are  the constituent masses $\mu_{[ud]}$,
$\mu_q$, eigenvalues $E_0$ in Eq. (\ref{shr}), and masses of
$[ud]^2\bar{q}$ states ($q=d,s$) for $J^P=\frac{1}{2}^\pm$.
$m_{[ud]}=0$, $m_d=0$, $m_s=0.17$ GeV}
\begin{center}
\begin{tabular}{|c|c|c|c|c|}
\hline & \(\mu_{[ud]}\) & \(\mu_q\) & \(E_0\) & \(M\) \\
\hline \([ud]^2\bar{s}~\frac{1}{2}^+\) & 0.467 & 0.470 & 1.561 & 2.115 \\
\hline \([ud]^2\bar{s}~\frac{1}{2}^-\) & 0.491 & 0.548 & 1.827 & 2.472 \\
\hline \([ud]^2\bar{d}~\frac{1}{2}^+\) & 0.452 & 0.452 & 1.584 & 2.051 \\
\hline \([ud]^2\bar{d}~\frac{1}{2}^-\) & 0.496 & 0.496 & 1.848 & 2.400 \\
\hline
\end{tabular}
\end{center}
\label{table_masses}
\end{table}

In conclusions, we have presented the first dynamical calculations
of the pentaquark masses within the Jaffe--Wilczek approximation.
Our predictions are universal in the sense that they use only two
parameters inherent to QCD: the string tension $\sigma$ and the
strong coupling constant $\alpha_s$. These results do not require
any additive mass shift used in constituent quark models. Our
predictions can be viewed as lower bounds for the masses of
pentaquark states. We showed that the $\Theta^+$ mass is much
higher than the observed one. This implies that $\Theta^+(1540)$
can not be explained in terms of the standard string interaction
between quarks in the diquark--diquark--antiquark picture. Note
that the chiral interactions (e.g. pion exchanges) are
automatically taken into account inside diquarks, while are not
possible between diquarks and strange antiquark.  Therefore a
drastic modification of the present results requires a completely
new dynamics, either the chiral soliton type, as in
\cite{Diakonov:1997}, or other quark clusters, like $q^4-\bar{q}$
\cite{Carlson} or $q^2\bar{s}-q^2$ \cite{Karliner}. \\

The authors are grateful to K.G.Boreskov for careful reading of
the manuscript and valuable suggestions. This work was supported
by RFBR grant $\#$ 03-02-17345 and PRF grant for leading
scientific schools $\#$ 1774.2003.2.


\begin{thebibliography}{99}
\bibitem{Nakano:2003}
T.~Nakano {\it et al.}  [LEPS Collaboration], Phys. Rev. Lett.
\textbf{91}, 012002 (2003); arXiv: hep-ex/0301020.
\bibitem{Barmin:2003}
V.~V.~Barmin {\it et al.}  [DIANA Collaboration], Yad. Fiz. {\bf
66} 1763 (2003), Phys. At. Nucl., {\bf 66} 1715 (2003); [arXiv:
hep-ex/0304040].
\bibitem{Stepanyan:2003}
S.~Stepanyan  [CLAS Collaboration], arXiv: hep-ex/0307018.
\bibitem{Kubarovsky}
V.~Kubarovsky, S.~Stepanyan [CLAS Collaboration], arXiv:
hep-ex/0307088.
\bibitem{Barth}
J. Barth {\it et al.} [SAPHIR Collaboration], arXiv:
hep-ex/0307083.
\bibitem{Dolgolenko}
A.~E.~Asratyan, A.~G.~Dolgolenko, and M.~A.~Kubantsev, arXiv:
hep-ex/0309042.
\bibitem{Diakonov:1997}
D.~Diakonov, V.~Petrov and M.~V.~Polyakov, Z. Phys.~ A {\bf 359},
305 (1997) [arXiv: hep-ph/9703373], M.Praszalovicz, arXiv:
 hep-ph/0308014.
\bibitem{Arndt}
R.A. Arndt, I.I. Strakovsky, and R.L. Workman, arXiv:
nucl-th/0308012.
\bibitem{Nussinov} S.~Nussinov, arXiv:
hep-ph/0307357.
\bibitem{capstick-page} S.Capstick, P.R.Page and W.Roberts, Phys. Lett. {\bf B570} (2003)
185-190; arXiv: hep-ph/0307019.
\bibitem{Jaffe:2003} R.L.~Jaffe, F.~Wilczek, arXiv: hep-ph/0307341
\bibitem{stancu}Fl. Stancu and  D. O. Riska, arXiv: hep-ph/0307010
\bibitem{sumrules} J.Sugiyama,
T Doi, M. Oka, arXiv: hep-ph/0309271
\bibitem{Fodor} Z.Fodor, arXiv: hep-lat/0309090
\bibitem{Sasaki} S.Sasaki, arXiv: hep-lat/0310014
\bibitem{anselmino} For a review see {\it e.g.} M.Anselmino {\it et al.}, Rev.
Mod. Phys. {\bf 65}, 1199 (1993) and references therein
\bibitem{lisbon} For a review see Yu.A.Simonov, in \emph{Proceedings of the XVII Autumn School
Lisboa, Portugal, 24 September - 4 October 1999}, edited by L.
Ferreira, P.Nogueira, and J. I. Silva-Marco (World Scientific,
Singapore, 2000), p.~60; [arXiv: hep-ph/9911237].
\bibitem{plekhanov} I.M.Narodetskii, A.N.Plekhanov, A.I.Veselov,
JETP Lett. {\bf 77} (2003) 58
\bibitem{CP-PACS} S.Aoki \textit{et al.} [CP-PACS Collaboration], arXiv: hep-lat/0206009
\bibitem{UKQCD} C.M.Maynard [UKQCD Collaboration] and D.G.Richard  [LHP Collaboration], arXiv: hep-lat/0209165
\bibitem{polyakov} A. M. Polyakov, \emph{Gauge Fields and Strings}
(Harwood Academic Publishers, 1987)
\bibitem{brin77} L. Brink, P. Di Vecchia, and P. Howe, Nucl. Phys.
{\bf B118}, 76 (1977).
\bibitem{simonov_self_energy}Yu.A.Simonov, Phys. Lett. \textbf{B515} (2001) 137
\bibitem{baryons} Yu.A.Simonov, Phys. Atom. Nucl. {\bf 66} (2003) 338-354; Yad. Fiz. {\bf 66} (2003)
363-380, arXiv: hep-ph/0205334
\bibitem{trusov} I.M.Narodetskii, M.A.Trusov, Phys. Atom. Nucl., to be published,
arXiv: hep-ph/0307131
\bibitem{fabre} M.Fabre de la Ripelle and Yu.A.Simonov, Ann. Phys. (N.Y.)
\textbf{212} (1991) 235
\bibitem{simonov_hyperspherical} Yu.A.Simonov, Yad. Fiz. \textbf{3}
(1966) 630, 1032
\bibitem{TMNS02} T.T.Takahashi, H.Matsufuru, Y.Nemoto and
H.Suganuma, Phys. Rev. Lett., {\bf 86} (2002) 18
\bibitem{capstick-isgur} S. Capstick and N. Isgur, Phys. Rev.  {\bf D34}, 2809
(1986).
\bibitem{Carlson} C.E.Carlson, C.D.Carone, H.J.Kwee, and
V.Nazaryan, arXiv: hep-ph/0307396
\bibitem{Karliner} M.Karliner and H.J.Lipkin, arXiv: hep-ph/0307243

\end{thebibliography}
\end{document}